\documentclass[aps,twocolumn,superscriptaddress,showpacs]{revtex4}
\usepackage{amsmath}
\usepackage{graphicx}
\usepackage{bm}
\usepackage{multirow}

\input{tcilatex}

\begin{document}

\title{Equilibrium and pre-equilibrium processes in the $^{\mathbf{55}}$Mn($%
^{\mathbf{6}}$Li,\emph{xp}) and $^{\mathbf{57}}$Fe($\boldsymbol{\alpha }$,%
\emph{xp}) reactions}
\author{A.V.~Voinov}
\email{voinov@ohio.edu}
\affiliation{Department of Physics and Astronomy, Ohio University, Athens, OH 45701, USA}
\author{S.M.~Grimes}
\affiliation{Department of Physics and Astronomy, Ohio University, Athens, OH 45701, USA}
\author{C.R.~Brune}
\affiliation{Department of Physics and Astronomy, Ohio University, Athens, OH 45701, USA}
\author{A.~B\"urger}
\affiliation{Department of Physics, University of Oslo, N-0316 Oslo, Norway}
\author{A.~G\"orgen}
\affiliation{Department of Physics, University of Oslo, N-0316 Oslo, Norway}
\author{M.~Guttormsen}
\affiliation{Department of Physics, University of Oslo, N-0316 Oslo, Norway}
\author{A.C.~Larsen}
\affiliation{Department of Physics, University of Oslo, N-0316 Oslo, Norway}
\author{T.N.~Massey}
\affiliation{Department of Physics and Astronomy, Ohio University, Athens, OH 45701, USA}
\author{S.~Siem}
\affiliation{Department of Physics, University of Oslo, N-0316 Oslo, Norway}
\author{C. Kalbach}
\affiliation{Physics Department, Duke University, Durham, NC 27708-0305, USA}

\begin{abstract}
Spectra of outgoing neutrons and protons from the $^{6}$Li+$^{55}$Mn
reaction and protons from the $\alpha $+$^{57}$Fe reaction have been
measured with beams of 15~MeV $^{6}$Li ions and 30~MeV $\alpha $ particles.
These reactions proceed through the same $^{61}$Ni nucleus at the same
excitation energy, thus allowing the difference in reaction mechanism to be
studied. It is shown that spectra from the first reaction measured at
backward angles are due to emission from a traditional compound nucleus
reaction, in which the intermediate nucleus has reached statistical
equilibrium; the spectra from the second reaction contain a significant
fraction of pre-equilibrium emission at all angles. Level density parameters
of the residual nucleus $^{60}$Co have been obtained from the first
reaction. Both emission spectra and angular distributions have been measured
for the second reaction. It was found that the pre-equilibrium component
exhibits a forward-peaked angular distribution, as expected, but with a
steeper slope than predicted and with an unusual slight rise at angles above
120$^{\circ }$. The backward-angle rise is explained qualitatively by the
dominance of the multi-step compound mechanism at backward angles.
\end{abstract}

\maketitle

\section{Introduction}

Understanding the pre-equilibrium reaction mechanism is one of the most challenging problems in
nuclear reaction physics. The term \textquotedblleft pre-equilibrium mechanism\textquotedblright\
refers to the process through which an incoming particle's energy is gradually redistributed among
more and more nuclear degrees of freedom. Energy equilibration can be described as a chain of
particle-hole excitations, where the particle and hole degrees of freedom are referred to
collectively as excitons. As the exciton number increases, some particle-hole pairs will
annihilate, and a state of full steady-state energy equilibrium is reached. An outgoing particle
can be emitted at any stage during the equilibration process, and this is usually referred to as
\textquotedblleft pre-equilibrium\textquotedblright\ emission, whereas particle emission occurring
after statistical equilibrium has been achieved is referred to either as equilibrium emission or
particle evaporation. Including the pre-equililbrium mechanism in calculations helps in
interpreting the observed enhancements of high energy particle emission compared to predictions
based solely on statistical evaporation at the equilibrium stage. In reality, however, there is no
sharp dividing line between the two stages, since equilibrium is approached very gradually.

The situation becomes complicated when calculating the energy and angular distributions of the
emitted particles. Sometimes, especially in studying the spectroscopy of low-lying energy states in
the residual nucleus, it is convenient to consider only direct reactions---those involving a single
target-projectile interaction. Some direct reaction mechanisms are inlcuded in common
pre-equilibrium models and others are not, nor is their universal agreement as to which mechanisms
are included in which models. In some applications it is useful to further divide the
pre-equilibrium component into two parts: multi-step direct (MSD) and multi-step compound (MSC).
Here, however, it is important to recognize that the equilibrium component represents additional
MSC cross section, even though it is typically calculated using traditional equilibrium models. The
MSD/MSC division of the cross section is discussed later in connection with the interpretation of
the experimental angular distributions.

Since the whole process is very complicated, there is still no unified
theory describing all aspects of it within the same framework. Instead,
there are several approaches describing each stage separately. For a review
of this topic, see Ref.\ \cite{Godioly}.

One factor hindering further development of the theories is the lack of
experimental data, especially double-differential cross sections of outgoing
particles. Most of the data are available for forward angles, and detailed
angular distributions for backward angles are often lacking or have poor
statistics. As a result, the most developed theories are those related to
the first stages of the equilibration process. These have the largest cross
sections and produce strongly asymmetric, forward-peaked angular
distributions. Angular distributions at backward angles are still not well
understood except for purely equilibrium reactions, for which they are
symmetric with respect to 90$^{\circ }$ in the center-of-mass system.

Although the angular distributions (pre-equilibrium and equilibrium) for
nucleon-induced reactions can be reasonably well described with the purely
phenomenological Kalbach systematics \cite{Kalbach}, there is a reported
problem for $\alpha $-particle-induced reactions, for which experimental
angular dependencies are often steeper than indicated by the systematics.

The more general problem directly related to the study of pre-equilibrium processes is to determine
the relative fractions of pre-equilibrium and equilibrium emission represented in experimental
energy spectra. Such a separation is usually model dependent and relies on input parameters used in
the Hauser-Feshbach compound nucleus reaction model. The most uncertain quantity is the level
density, which strongly affects the shape of particle evaporation spectra. Therefore, the level
density appears to be crucial for the separation of experimental spectra into their pre-equilibrium
and equilibrium components.

In this work we study inclusive proton spectra from the $^{55}$Mn($^{6}$Li,$%
xp$) and $^{57}$Fe($\alpha $,$xp$) reactions experimentally. These reactions proceed through the
same $^{61}$Ni intermediate nucleus, and our incident energies are chosen to produce it with very
nearly the same excitation energy. We will show that the spectrum from the first reaction measured
at backward angles can be considered entirely due to the compound nucleus (or equilibrium
evaporation) mechanism, whereas a large fraction of the ($\alpha $,$xp$) cross section is known to
be due to pre-equilibrium processes. Therefore, by comparing proton spectra from these two
reactions, the pre-equilibrium fraction in the latter reaction can be obtained in a nearly model
independent way. In addition, the level density of the residual nucleus $^{60}$Co can be obtained
from the proton evaporation spectrum of the lithium-induced reaction, and our understanding of this
reaction can be verified by simultaneously reproducing the ($^{6}$Li,$xn$) reaction using the same
model calculations.

\section{Experiments}

\subsection{The $^{\mathbf{55}}$Mn($^{\mathbf{6}}$Li,\emph{xp}) and $^{%
\mathbf{55}}$Mn($^{\mathbf{6}}$Li,\emph{xn}) reactions}

The inclusive proton and neutron spectra from the $^{6}$Li+$^{55}$Mn reaction have been measured
with a 15 MeV $^{6}$Li beam from the tandem accelerator of the Edwards Accelerator Laboratory, at
Ohio University. The protons have been registered with a $\Delta E$-$E$ telescope consisting two
silicon detectors. Their thicknesses were 150 $\mu $m for the $\Delta E$ detector and 3500 $\mu $m
for the $E$ detector. The latter was able to stop protons with energies up to 28~MeV, which was
sufficient to measure the proton spectrum up to its maximum energy of around 25~MeV. The telescope
was placed at a laboratory angle of 150$^{\circ }$, to reduce or eliminate the contribution from
non-equilibrium mechanisms. Because of the stopping power of the $\Delta E$ detector, only protons
with energies above about 7~MeV were registered. The low energy portion of the proton spectrum,
with proton energies up to about 10 MeV, was measured with the charged-particle spectrometer
\cite{Voinov}, in which the energy and the particle type were determined by the energy deposited in
1500 $\mu $m Si detectors and the flight time for the 2~m flight paths between the target and the
detector. The measurement was carried out at 157.5$^{\circ }.$ The absolute cross section was
obtained from the known target thickness, the integrated beam current, and the solid angle of the
detectors. The target for all of the $^{55}$Mn measurements was a 1 $\mu $m thick manganese foil
coated with a thin (10~nm) carbon foil. The background caused by the carbon layer has been
determined from a separate experiment on a carbon target and found to be negligible.

The neutron spectrum was measured at 125$^{\circ }$ and 140$^{\circ }$ with the neutron
time-of-flight spectrometer of the Edwards Accelerator Laboratory. The 5~m flight path and NE-213
liquid scintillators were used to determine the energy of the outgoing neutrons. Detector
efficiencies were determined with the calibrated neutron spectrum from the $^{27}$Al($d$,$n$)
reaction measured with 7.5~MeV deuterons \cite{Aleff}. The spectra at the two angles were identical
to within their error bars and were averaged for comparison with model calculations.

\subsection{The $^{\mathbf{57}}$Fe($\boldsymbol{\protect\alpha }$,\emph{xp})
reaction}

Protons from the 30 MeV $\alpha $-particle-induced reactions on $^{57}$Fe
have been measured with the $\Delta E$-$E$ technique at the cyclotron
laboratory of the University of Oslo. The $\Delta E$-$E$ telescope consisted
of a 200~$\mu $m thick $\Delta E$ Si detector and a 5000~$\mu $m thick
Si(Li) detector. It was rotated around the target to measure the angular
distributions. Spectra were measured at nine angles: 30$^{\circ }$, 45$%
^{\circ }$, 70$^{\circ }$, 90$^{\circ }$, 104$^{\circ }$, 120$^{\circ }$, 135%
$^{\circ }$, 150$^{\circ }$, and 160$^{\circ }$ in the laboratory system. A
separate 1500~$\mu $m thick silicon detector was placed at 45$^{\circ }$ to
monitor the beam current by measuring elastically scattered $\alpha $
particles. The 1.7\textrm{~}mg/cm$^{2}$ thick $^{57}$Fe target was enriched
to 95\%.

\begin{figure}[t]
\includegraphics[width=9cm]{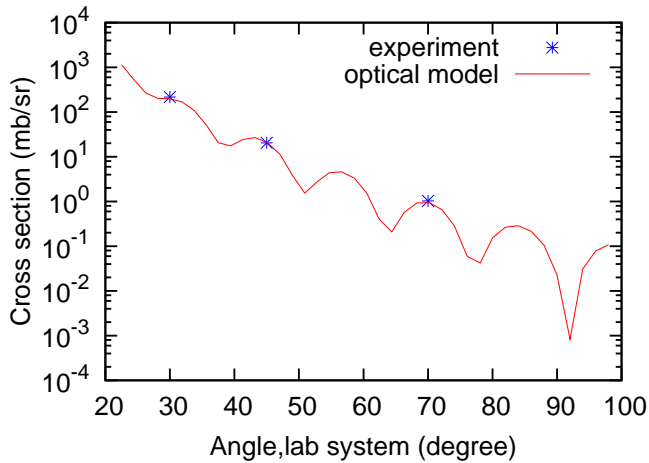} .
\caption{The angular dependence of elastically scattered $\protect\alpha $
particles. Points show the experimental data scaled to the calculation.
Error bars due to counting statistics are less than the size of the points.
The line show the results of calculations with the optical model potential
N-9401 from the RIPL-2 database \protect\cite{RIPL}}
\label{fig:fig1}
\end{figure}

In order to obtain absolute cross sections, the angular dependence of the
elastically scattered $\alpha $ particles was measured and scaled to the
results of optical model calculations. Only three data points from forward
angles were used for the scaling, because at backward angles inelastic $%
\alpha $ particles populating the first excited level of $^{57}$Fe, at
14.4~keV, might contaminate the elastic-scattering peak. Different optical
model parameters taken from the RIPL-2 compilation \cite{RIPL} were tested.
The best parameters were found to be under number 9401 in the compilation.
Figure\ \ref{fig:fig1} presents the scaled experimental points along with
the results of the optical model calculations. The uncertainty in this
scaling is mostly determined by the average deviation of the scaled
experimental points from the calculated ones, which is about 8\% in our
case. Uncertainties due to counting statistics do not exceed 1\%.

\section{Analysis of experimental spectra}

The experimental proton and neutron spectra from both reaction systems were converted to the
center-of-mass system for comparison with calculations. The usual assumption that all particles
were emitted from the fused target+projectile nucleus was made.

\subsection{The $^{\mathbf{55}}$Mn($^{\mathbf{6}}$Li,\emph{xp}) and $^{%
\mathbf{55}}$Mn($^{\mathbf{6}}$Li,\emph{xn}) reactions}

\begin{figure*}[tbp]
\includegraphics[width=17cm]{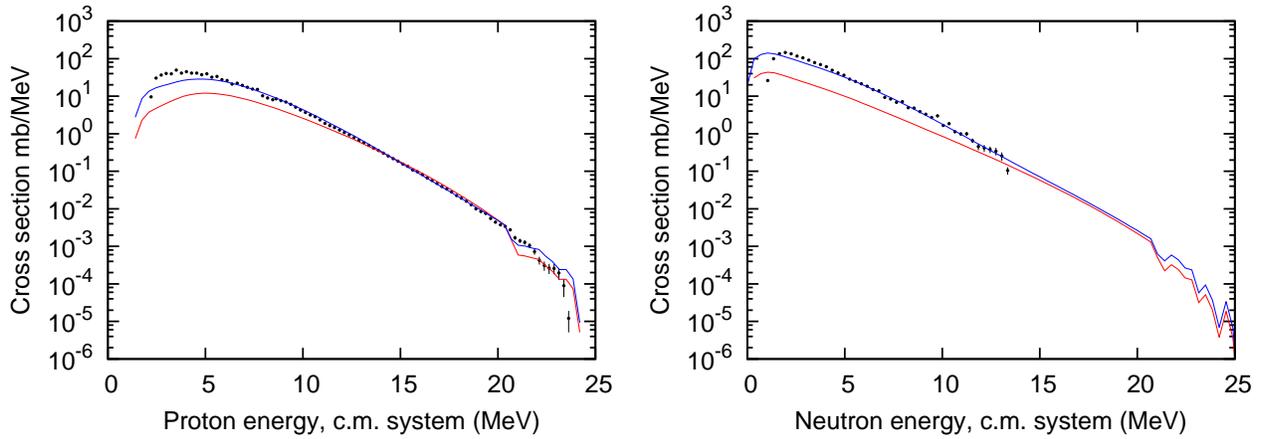}
\caption{Proton and neutron evaporation spectra from the $^{6}$Li+$^{55}$Mn
reaction. The points are experimental data taken at 157$^{\circ }$ for the
lower proton energies and 150$^{\circ }$ for higher proton energies. The
lines are the result of \textsc{empire} calculations with Gilbert-Cameron
(top curve) and HF-BCS (bottom curve) input level density models.
Experimental cross sections are multiplied by 4$\protect\pi $~sr to compare
with the calculated, angle-integrated spectrum.}
\label{fig:fig2}
\end{figure*}

The proton and neutron spectra from the $^{6}$Li+$^{55}$Mn reaction measured
at backward angles have been analyzed with the \textsc{empire} computer code
\cite{Empire} using the Hauser-Feshbach compound-nucleus-reaction model \cite%
{hauser} (see Fig.\ \ref{fig:fig2}). No pre-equilibrium emission was allowed
in the calculations, but sequential equilibrium emission of more than one
particle was considered. Calculations were performed with two sets of model
level densities. One was the Gilbert-Cameron composite formula \cite{GC1},
which uses a constant-temperature dependence at low energies and a Fermi-gas
dependence at higher excitation energies, while the other was the
microscopic model based on a statistical approach using the HF-BCS model
\cite{Goriely}. Parameter systematics for the Gilbert-Cameron formula were
taken from Ref.\ \cite{Arthur}. The Fermi-gas parameter $a$ in the \textsc{%
empire} code is assumed to be energy dependent according to the Ignatyuk
formula \cite{Ignatuk}, which takes into account shell effects. The level
density parameters for the residual $^{60}$Co nucleus are as follows: the
nuclear temperature $T$ is 1.35~MeV, the adjustable energy shift $E_{0}$ is
-3.39~MeV, the pairing energy $\Delta $ is 0.0~MeV, the asymptotic parameter
$\tilde{a}$ is 7.95~MeV$^{-1}$ and the parameter $\gamma $ in the Ignatyuk
formula is -0.054. We also tested the functional form of the original
Fermi-gas (Bethe) formula \cite{Bethe} (without the constant temperature
part) and found it to be totally consistent with the HF-BCS level densities
for the residual nuclei populated in this reaction.
We found that the experimental spectra are best reproduced with the
Gilbert-Cameron formula, which includes the constant-temperature energy
dependence at low excitation energies. Models using only the Fermi-gas
energy dependence do not reproduce the shape of the evaporation spectra.
This conclusion is in agreement with our previous results \cite{CT} and the
results of Ref.\ \cite{sherr}, where the authors showed the importance of
the constant-temperature level density in describing evaporation spectra in
the $A=50$ to 60 mass range.

\begin{figure}[tbp]
\includegraphics[width=9cm]{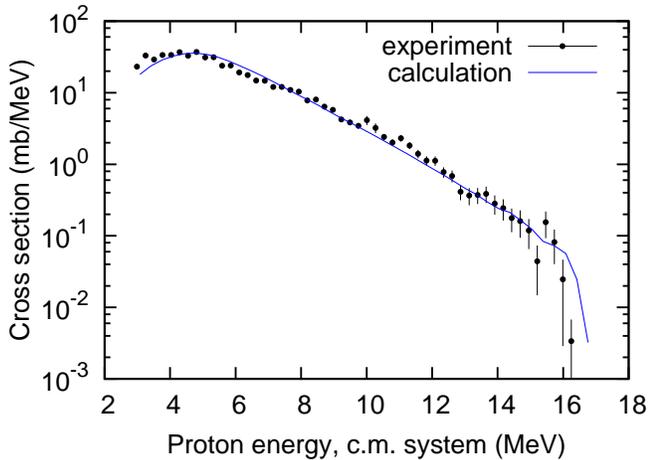}
\caption{The proton evaporation spectrum from the $^{3}$He+$^{58}$Fe
reaction. The points show the 157$^{\circ }$ data from Ref.\ \protect\cite%
{Voinov} normalized to 4$\protect\pi $ sr, and the line shows the
angle-integrated results of \textsc{empire} calculations using the
Gilbert-Cameron level density model with the parameters from Ref.\
\protect\cite{Arthur}.}
\label{fig:fig3}
\end{figure}

We also compared results from the $^{55}$Mn($^{6}$Li,$xp$) proton spectrum
with information from the $^{58}$Fe($^{3}$He,$xp$) reaction. The latter was
measured with a 10~MeV $^{3}$He beam, also at the Edwards Accelerator
Laboratory \cite{Voinov}. In Fig.\ \ref{fig:fig3} we compare the
experimental proton spectrum from the $^{3}$He-induced reaction with
Hauser-Feshbach calculations using the same Gilbert-Cameron level densities
that were used to calculate the spectrum from the lithium-induced reaction
(see Fig.\ \ref{fig:fig2}). The conclusion is that the Gilbert-Cameron level
density is able to reproduce the backward-angle proton spectra from both
reactions. 

The fact that proton spectra from both reactions are described with the same
input level density function, even though the excitation energies in the
intermediate nucleus were different, indicates the independence of the
proton emission spectra on the type of projectile. We thereby conclude that
at these incident energies, the dominant reaction mechanism at backward
angles in both reactions and for all proton emission energies is particle
emission from an energy-equilibrated compound nucleus.

\subsection{The $^{\mathbf{57}}$Fe($\boldsymbol{\protect\alpha }$,\emph{xp})
reaction}

Proton spectra from the $^{57}$Fe($\alpha $,$p$) reaction were measured at
each of the nine experimental angles. The results are presented in Fig.\ \ref%
{fig:fig8}, and the angle-integrated energy spectrum is shown in Fig.\ \ref%
{fig:fig4}. An unusual feature of the data is the concentration of
measurements at angles in the backward hemisphere with very good statistics
(especially at 160 degrees) even at the higher emission energies, where the
cross section is over two orders of magnitude lower than in the evaporation
peak. This facilitates the study of the angular distributions at backward
angles.
\begin{figure}[tbp]
\includegraphics[width=9cm]{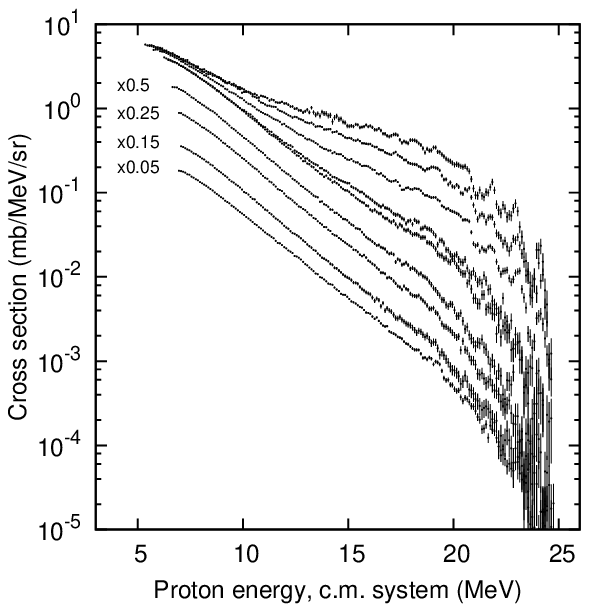}
\caption{Proton experimental spectra from the $\protect\alpha $+$^{57}$Fe
reaction measured at laboratory angles of 30, 45, 70, 90, 104, 120, 135,
150, and 160 degrees (from top to the bottom). The four backward-angle
spectra are scaled down for better visualization.}
\label{fig:fig8}
\end{figure}
\begin{figure}[tbp]
\includegraphics[width=9cm]{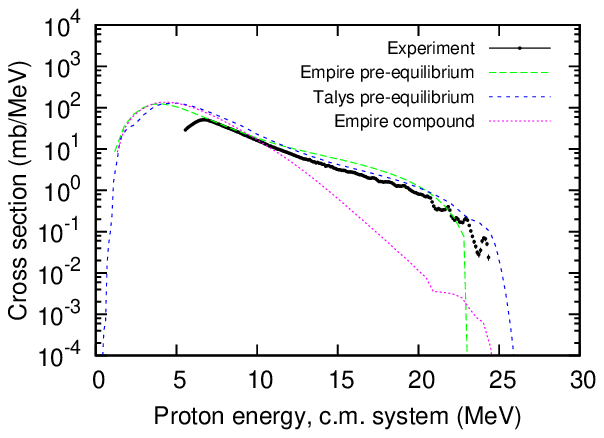}
\caption{The angle-integrated proton spectrum from the $\protect\alpha $+$%
^{57}$Fe reaction. The points show the experimental data, while the lines
are the results of calulations using the \textsc{empire} and \textsc{talys}
codes. The curves labeled \textquotedblleft
pre-equilibrium\textquotedblright\ also include contributions from the
equilibrium or compound component.}
\label{fig:fig4}
\end{figure}
The calculated spectra shown in Fig.\ \ref{fig:fig4} were obtained using the
\textsc{talys} \cite{Talys} and \textsc{empire} \cite{Empire} computer
codes. While both codes do well at reproducing the overall energy
distribution of the emitted particles, \textsc{talys} uses the more
sophisticated two-component exciton pre-equilibrium model and gives better
agreement with the data, especially at the highest emission energies.
Therefore we will use \textsc{talys} calculations to further analyze the
proton angular distributions. The small overall excess (25\%) in the
calculated cross sections relative to the experimental ones is well within
the level of agreement one might expect from statistical models using a
global input set. It can, in part, be explained by inadequacies in the level
density parameters for the residual nuclei populated by the competing proton
and neutron emission channels and/or by small contributions from direct
processes not taken into account in the model calculations. Emission from
the equilibrated compound nucleus was calculated with the input level
density obtained from the $^{55}$Mn($^{6}$Li,$xp$) reaction, which thus
forms a useful complement to the $^{57}$Fe($\alpha $,$xp$) reaction.

\begin{figure}[tbp]
\includegraphics[width=9cm]{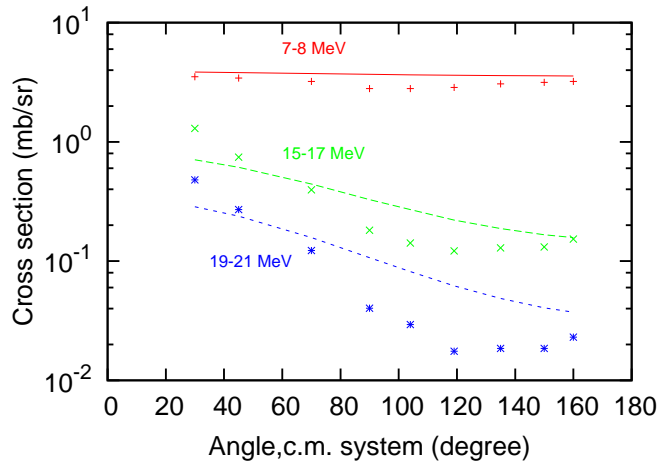}
\caption{Angular distribution of outgoing protons from different energy
intervals. The points show the experimental results from the $^{57}$Fe($%
\protect\alpha $,$xp$) reaction, and the lines are from the calculations
with the code \textsc{talys}.}
\label{fig:fig5}
\end{figure}

The angular distributions for several energy bins of outgoing protons are presented in Fig.\
\ref{fig:fig5}, along with the results of \textsc{talys} calculations. These calculations assume
that the equilibrium component, calculated with the Hauser-Feshbach-model part of the code, is
emitted isotropically. This is often a good approximation, but the emission is better described as
being symmetric about 90$^{\circ }$ in the center-of-mass system, typically with a small dip or
minimum at that angle. For the pre-equilibrium component, \textsc{talys} uses the phenomenological
Kalbach angular-distribution systematics \cite{Kalbach}. These systematics are based on
experimental results for a wide variety of reaction channels. Figure \ref{fig:fig5} shows that for
low-energy outgoing protons, the experimental angular distributions are flat, and their
near-symmetry about 90 degrees indicates that evaporation from the compound nucleus is the dominant
reaction mechanism. Because of the assumption of isotropic emission for the equilibrium component,
the calculations show much less of a dip around 90$^{\circ }$ than the data but otherwise give
reasonable agreement with them. For more energetic protons, the equilibrium component decreases
rapidly in intensity, and the angular distributions exhibit the forward-peaked behavior
characteristic of pre-equilibrium processes. These angular distributions are not well reproduced by
the \textsc{talys} results, which are much less forward-peaked than the data and fail to show the
flat or slightly increasing cross section at angles above 120$^{\circ }$.

\begin{figure}[tbp]
\includegraphics[width=9cm]{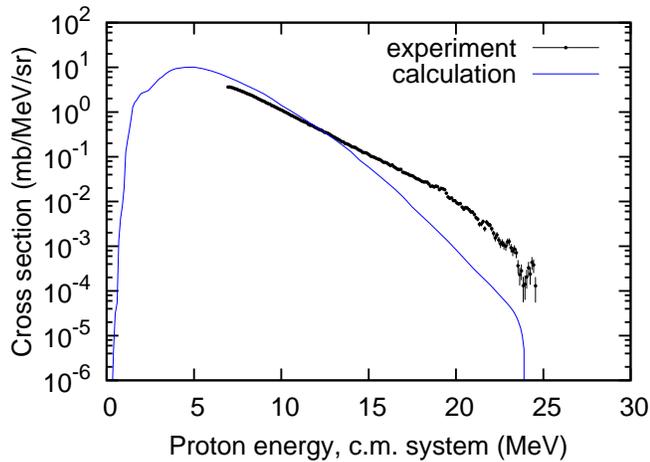}
\caption{Energy spectrum of outgoing protons from the $^{57}$Fe($\protect%
\alpha $,$xp$) reaction measured at a laboratory angle of 160$^{0}$. The
points show the data, and the line is the compound nucleus reaction
component calculated in the \textsc{talys} code.}
\label{fig:fig6}
\end{figure}

\begin{figure}[tbp]
\includegraphics[width=9cm]{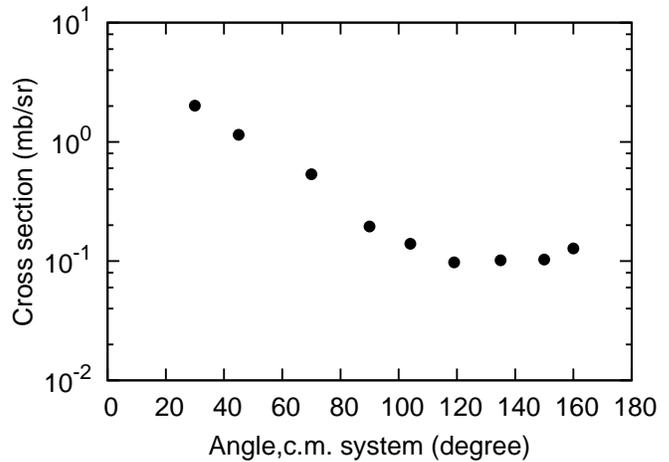}
\caption{Experimental angular distribution of pre-equilibrium outgoing
protons with energies greater than 16~MeV from the $^{57}$Fe($\protect\alpha
$,$p$) reaction.}
\label{fig:fig7}
\end{figure}

To better understand this discrepancy between experiment and the Kalbach angular-distribution
systematics used in the model calculation, it is useful to try to divide the measured cross section
into its equilibrium and pre-equilibrium components. Fortunately, this is possible because the
equilibrium component can be calculated based on the data from the $^{55}$%
Mn($^{7}$Li,$xp$) reaction. Figure \ref{fig:fig6} compares the experimental 160$^{\circ }$ proton
spectrum with the equilibrium component calculated by \textsc{talys}. By subtracting the
equilibrium component from the experimental spectra at each angle, the double-differential cross
sections and thus the angular distribution of the experimental pre-equilibrium cross section can be
determined. The calculated evaporation component was normalized downward by about 20\% to reproduce
the data at low emission energies before doing the subtraction. Figure \ref{fig:fig7} shows the
angular distribution of the resulting pre-equilibrium component from the $^{57}$Fe($\alpha $,$xp$)
reaction integrated over emission energies from 16 to 25 MeV, where pre-equilibrium emission is
dominant and the subtraction is most accurate. If the pre-equilibrium component were pure MSD, as
is assumed in the model calculations, then this angular distribution should continue to decrease
with increasing angle. Instead it is again flat above 120$^{\circ }$, with a small increase at
160$^{\circ }$. The observed cross section at backward angles therefore implies that the
pre-equilibrium component is not pure MSD but contains significant MSC pre-equilibrium cross
section.  This pre-equilibrium MSC cross section, because its angular distribution is symmetric
about 90$^{\circ}$, produces the observed rise in cross section at large angles in the figure. In
this regard, it is interesting to note that the data from Ref.\ \cite{Alera} showed a similar
backward-angle rise in cross section for outgoing protons in the $^{56}$Fe($%
\alpha $,$xp$) reaction at an incident energy of 23 MeV. All of this points
to the need to reexamine the angular distribution systematics for the ($%
\alpha $,$xp$) reaction channel, recognizing that some of the pre-equilibrium cross section will be
multi-step compound in nature.

\section{Implications for the Kalbach systematics}

The Kalbach angular distribution systematics \cite{Kalbach} utilize the division of the cross
section into its multi-step direct (MSD) and multi-step compound (MSC) parts. Multi-step direct
emission is defined as particle emission that occurs when the equilibrating nucleus has passed
through a sequence of states all of which have at least one
particle-degree-of-freedom in an unbound single-particle state \cite%
{Feshbach1}. It is made up of the direct and much of the pre-equilibrium cross section, and it
exhibits a forward-peaked angular distribution described by an exponential in $\cos \theta $. Once
the system passes through at least one configuration in which all of the particle
degrees-of-freedom are in bound single-particle states, \textquotedblleft memory\textquotedblright\
of the direction of motion of the projectile is assumed to be lost and the mechanism is termed
multi-step compound. Subsequent particle emission is symmetric about 90$^{\circ }$ in the
center-of-mass and contains exponentials in both $\cos \theta $ and $-\cos \theta $. The MSC cross
section is composed of the rest of the pre-equilibrium component plus the equilibrium cross
section. Thus, the systematics, when applied to equilibrium emision, avoid the assumption of
isotropy used in \textsc{talys} and \textsc{empire} for that component.

The angular distribution of the double-differential cross section has the
form
\begin{align}
\frac{d^{2}\sigma }{d\Omega d\epsilon _{b}}& =\frac{1}{4\pi }\,\frac{d\sigma
}{d\epsilon _{b}}\,\frac{1}{e^{a_{\text{\textrm{ang}}}}-e^{-a_{\text{\textrm{%
ang}}}}}\left[ (1+f_{\mathrm{msd}})e^{a_{\text{\textrm{ang}}}\cos \theta
}\right.   \notag \\
& +\left. (1-f_{\mathrm{msd}})e^{-a_{\text{\textrm{ang}}}\cos \theta }\right]
\label{AngDisEq}
\end{align}%
where $\epsilon _{b}$ is the channel energy for the exit channel in the
reaction, $f_{\mathrm{msd}}$ is the fraction of the cross section that is
MSD, and $a_{\mathrm{ang}}$ is the angular distribution \textquotedblleft
slope parameter,\textquotedblright\ used for both the forward-peaked MSD and
symmetric MSC components. A slope parameter of zero would yield an isotropic
angular distribution.\ The Kalbach systematics define the slope parameter as
a function of the energies of the incident and emitted particles, and its
values have been set phenomenologically, based on comparisons with a broad
database. 

In the later stages of developing these systematics and in their implementation in \textsc{talys},
it was assumed that the pre-equilibrium component is purely MSD. The comparison of experimental
angular distributions with the systematics at 16 and 20 MeV (see Fig.\ \ref%
{fig:fig5}), however, suggests that this assumption was inadequate and that\ the systematics need
to be revised to be able to describe the data, especially at the higher emission energies where
neither the amount of forward peaking nor the rise at backward angles is reproduced.

\begin{figure}[tbh]
\begin{center}
\includegraphics[
trim=0.000000in 0.000000in -0.003975in 0.000000in, height=3.1298in, width=3.039in ]{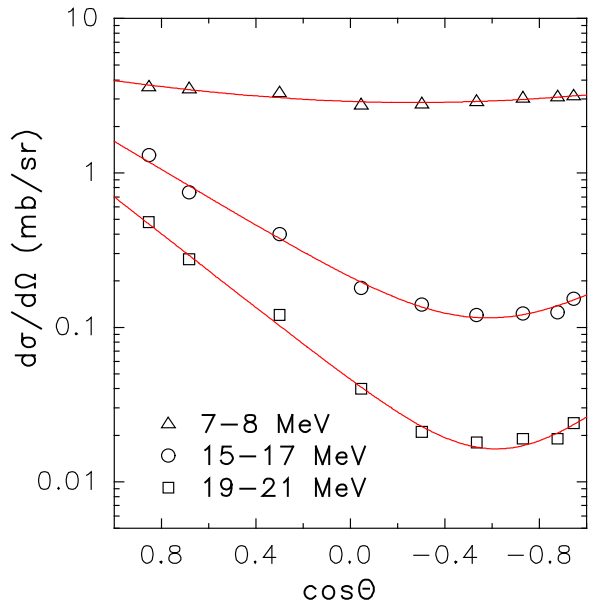}
\end{center}
\caption{Comparison between the experimental angular distributions for the $%
^{57}$Fe($\protect\alpha $,$xp$) reaction and the curves obtained by fitting
using Eq.\ (\protect\ref{AngDisEq}). The results are displayed as a function
of $\cos \protect\theta $ in order to show the exponential fall off of the
cross section at forward angles at 16 and 20 MeV.}
\label{AngDisFit}
\end{figure}

To quantify the problem, the experimental angular distributions from Fig.\ %
\ref{fig:fig5} were fit using Eq.\ (\ref{AngDisEq}) while varying the
parameters $d\sigma /d\epsilon $, $f_{\mathrm{msd}}$, and $a_{\mathrm{ang}}$%
. The results, shown in Fig.\ \ref{AngDisFit}, demonstrate that the general form of the equation
successfully reproduces the data. However the
fitting process yields parameter values, summarized in Table \ref%
{Parameters}, which are quite different from the values from Ref.\ \cite%
{Kalbach} used in \textsc{talys}. As expected, the slope parameters at 16 and 20 MeV are much
larger than indicated by the Kalbach systematics. In addition, the values for $f_{\mathrm{msd}}$ at
these energies are lower than $f_{\mathrm{pre}}$, the fraction of the
cross section due to pre-equilibrium emission, obtained from \textsc{talys}%
. This confirms that there is likely to be an MSC contribution to the pre-equilibrium component.
The fitted curves show that even a small amount of MSC cross section (6\% to 7\% at 20 MeV) can
cause a significant rise in the cross section at backward angles when a large slope parameter
causes the main, MSD part of the angular distribution to fall off rapidly with increasing angle.
The existence of multi-step compound pre-equilibrium emission is known, but for nucleon-induced
reactions it is most often concentrated in the region of the evaporation peak \cite{Kalbach1}. The
enhanced MSC emission for high energy protons from $\alpha $-induced reactions is a new result
which needs to be understood.

\begin{table}[th]
\caption{Comparison of parameters from fitting the experimental angular
distributions with the parameters used in \textsc{talys}.}%
\begin{tabular}{cccccc}
\hline\hline
Proton & $a_{\text{ang}}$ & $a_{\text{ang}}$ & sys/fit & $f_{\text{pre}}$ & $%
f_{\text{msd}}$ \\
energy & (sys.) & (fit) &  & (calc.) & (fit) \\ \hline
7.5 MeV & 0.62 & 0.67(7) & 1.1 & 0.05 & 0.18(3) \\
16 MeV & 0.98 & 2.10(8) & 2.1 & 0.90 & 0.84(1) \\
20 MeV & 1.15 & 2.75(9) & 2.4 & 0.99 & 0.935(6) \\ \hline\hline
\end{tabular}%
\label{Parameters}
\end{table}

\section{Discussion}

Physically, the results of the fitting process imply, for instance, that for
6\% to 7\% of the pre-equilibrium cross section for 20 MeV proton emission,
the system has passed through one or more configurations where all the
proton particle-degrees-of-freedom are in single particle states \emph{below}
the proton binding energy and that one such degree of freedom then gets
promoted to a single particle state 20 MeV \emph{above} the proton binding
energy, through one or more two-body interactions. This kind of MSC
pre-equilibrium emission is clearly possible, because we see equilibrium
evaporation of such energetic particles, but detailed calculations in a
single, unified framework will be needed to see if pre-equilibrium MSC
emission can account for as much cross section as is indicated by fits to
the data.

The probable presence of MSC pre-equilibrium cross section at such high emission energies (relative
to the spectral endpoint) suggests that too much weight was given to backward-angle data
(frequently with small cross sections and large error bars) in determining the slope parameters in
the Kalbach systematics for $\alpha $-particle-induced reactions, because that analysis assumed
that pre-equilibrium particle emission was pure MSD. For $\alpha $-particle induced reactions, the
observed excess of forward-angle experimental cross section that was not reproduced by the
resulting systematics was found in Ref.\ \cite{Kalbach} to be qualitatively consistent with the
observation of projectile-breakup fragments traveling with roughly the projectile velocity.
However, a new preliminary model of light-projectile breakup \cite{Kalbach3} suggests that the
breakup peaks should be less intense and have a narrower energy distribution than the excess cross
section observed in Ref.\ \cite{Kalbach}.
This is again consistent with the need to reconsider the systematics for $%
\alpha $-particle-induced reactions, putting more weight on intermediate and
forward angles and less on the very backward angles.

\section{Conclusion}

The proton spectra from both the $^{55}$Mn($^{6}$Li,$xp$) and $^{57}$Fe($%
\alpha $,$xn$) reactions have been measured and compared with calculations
performed using the exciton pre-equilibrium model and a Hauser-Feshbach
evaporation model.

The backward-angle spectrum from the first reaction allows the level density
parameters and the equilibrium component to be well determined, because the
measured cross section appears to be almost exclusively due to particle
evaporation from an energy-equilibrated system. Here it is found that the
Gilbert-Cameron level density model with the input parameters of Ref.\ \cite%
{Arthur} works well. This was confirmed by comparisons with earlier results
on the $^{58}$Fe($^{3}$He,$xp$)$^{60}$Co reaction at 10 MeV.

The angular distributions from the $^{57}$Fe($\alpha $,$xp$) reaction show two important effects.
First, for emission energies above the main evaporation peak, the experimental cross section
exhibits a significantly more rapid decrease with increasing angle than the calculated curves. This
suggests that the Kalbach systematics for the ($\alpha $,$xp$) channel and perhaps for all
reactions with incident $\alpha $-particles need to be revised. Second, the excellent data
statistics in the backward-angle spectra (in particular the 160$^{\circ }$ spectrum) show a
distinct flattening of the angular distributions and even a small rise in cross section at angles
above 120$^{\circ }$ for the higher emission energies. Using the level-density parameters
determined from the ($^{6}$Li,$xp$) reaction, it is shown that most of the cross section in this
domain is not due to equilibrium evaporation of protons, so there are pre-equilibrium processes
that exhibit a backward rise in cross section. This suggests that there is more multi-step compound
cross section in the pre-equilibrium component at these energies than one might na\"{\i}vely
expect, especially given the predicted predominance of direct nucleon-transfer (stripping)
reactions at proton energies of 16 and 20 MeV. This observation can help
guide a re-evaluation of the angular distribution systematics for incident $%
\alpha $ particles.

\section{Acknowledgments}

We greatly acknowledge the help of D. Jacobs, D. Carter, J.O'Donnell for running the Edwards
accelerator and for the computer and electronics support. We are also very grateful to E.A. Olsen
and A. Semchenkov for running the Oslo Cyclotron. Financial support from the Research Council of
Norway and U.S. Department of Energy grants numbers DE-FG52-06NA26187, DE-FG02-88ER40387, \ and
DE-FG02-97ER41033 (Duke University, for C.K.) is greatly appreciated.


\end{document}